\documentclass[preprint, 5p, twocolumn]{elsarticle}
\journal{Physics Letters A}
\usepackage[colorlinks=true, citecolor=red, urlcolor=blue ]{hyperref}
\usepackage{epsfig,newlfont,bm,subfigure,palatino,mathtools,braket,times,soul,setstack}
\usepackage{color}
\usepackage[utf8]{inputenc}
\usepackage[sc,osf]{mathpazo}
\usepackage[normalem]{ulem}
\usepackage[T1]{fontenc}
\usepackage{multirow}
\usepackage{graphics}
\usepackage{graphicx}
\usepackage{latexsym}
\usepackage{amssymb}
\usepackage{color}
\usepackage{graphics,epstopdf}
\usepackage{capt-of}
\usepackage{lipsum}
\usepackage{adjustbox}
\usepackage[table,xcdraw]{xcolor}
\usepackage{physics}
\usepackage{ragged2e}
\usepackage{comment}
\usepackage{widetext}
\usepackage[font=small,labelfont=bf,justification=justified,singlelinecheck=false]{caption}

\newcolumntype{P}[1]{>{\centering\arraybackslash}p{#1}}

\begin{document}

\title{ Coherence measure of ensembles with nonlocality  without entanglement }

\author{Ayan Patra$^{1}$, Shiladitya Mal$^{1, 2, 3,4}$, Aditi Sen(De)$^{1}$}
	
	\address{$^1$ Harish-Chandra Research Institute,  A CI of Homi Bhabha National Institute, Chhatnag Road, Jhunsi, Prayagraj - $211019$, India}
	\address{$^2$ Department of Physics and Center for Quantum Frontiers of Research and Technology (QFort),
National Cheng Kung University, Tainan 701, Taiwan}
\address{$^3$ Physics Division, National Center for Theoretical Sciences, Taipei 10617, Taiwan}
\address{$^4$ Centre for Quantum Science and Technology, Chennai Institute of Technology, Chennai 600069, India
	}
	\begin{abstract}
 \textcolor{black}{Irreversibility between preparation and discrimination processes is manifested in the  indistinguishability  of orthogonal product states via local operations and classical communication (LOCC). Characterizing quantum properties for sets of states according to their local distinguishing property is one of the avenues to explain the surprising results obtained in the LOCC indistinguishability domain. We propose a measure based on the \(l_1\) norm of coherence to quantitatively assess the quantumness of ensembles composed of orthogonal product states.  Furthermore, to establish a hierarchy among different product ensembles, we establish a relationship between the coherence-based measure of an ensemble and the optimal success probability of distinguishing states within the ensemble using LOCC, constrained by a limited amount of classical communication and projective measurements, in the framework of minimum error state discrimination.}   
	\end{abstract}
\maketitle
\section{Introduction}
\label{sec:intro}
One of the fundamental tasks in  quantum mechanics is to detect quantum states that are given from a known ensemble. Under globally allowed operations, sets of orthogonal quantum states can  always be distinguished while for nonorthogonal ensembles, the useful upper bound on the accessible information, quantifying the maximum amount of information extractable from the ensemble is known as the Holevo quantity \cite{HOLEVO-Holevobound}. 

On the other hand,  if quantum information is encoded into the composite system and subsystems of it are sent to  spatially separated observers,   a set of globally orthogonal states, in general,  cannot be distinguished under a set of allowed operations,  local operations assisted by classical communications (LOCC)  which is a strict subset of global operations \cite{PERES-locc}. Initially, it was thought that  entanglement which cannot be created by LOCC is responsible for 
local indistinguishability. However, such intuitive understandings turn out to be false on several occasions.  One of the surprising results in this direction is the discovery of a set consisting of nine orthogonal product states of two qutrits, which cannot be distinguished perfectly by LOCC - known as `nonlocality without entanglement' \cite{Bennett99}. 
 In a similar spirit, unextendible product qudit basis (UPB) have been discovered  \cite{Bennett99a,DIVINCENZO-upb}, which are also LOCC indistinguishable and they provide a systematic way of constructing bound entangled states \cite{HORODECKI-Boundentangledstate,Bennett99a,DIVINCENZO-upb}. Further investigations in this direction were carried out which found   several complete and incomplete LOCC indistinguishable product ensembles  as well as \textcolor{black}{distinguishable product ensembles with a finite rounds of LOCC protocol} \cite{VAIDMAN-distprodst,RINALDIS-distprodst,NATHANSON-distprodst,WANG-distprodst,ZHANG-distprodst,SARONATH2-nonlocw/oent}.  On the other hand, two orthogonal states are shown to be always distinguishable via LOCC irrespective of their entanglement content \cite{WAlGATE&HARDY-loccdist}. Moreover, it was exhibited in two qutrits that decreasing average entanglement from ensembles can increase local indistinguishability, a phenomenon known as ``more nonlocality with less entanglement" \cite{ASD-morenonlessent}. All the results strongly indicate that there are other quantum characteristics in ensembles different than average entanglement content which can be responsible for LOCC indistinguishability. To characterize this, several measures of quantum correlations beyond entanglement are introduced \cite{Horodecki_PRA_2005,Ollivier_PRL_2001,Oppenheim_PRL_2003,Henderson_JOPA_2001,Bera_ROPP_2017}.

Over the years, the studies of local indistinguishability are performed into two distinct directions -- on one hand, several counter-intuitive examples of ensembles that are LOCC indistinguishable are reported, while on the other hand, there are few attempts to quantify quantumness in the ensembles which can capture the difficulties in local distinguishing \cite{HORODECKI-loccdist, HORODECKI-distprotocol, HORODECKI-quantumcorrelation, ASD-lowerboundonlocaccinfo, feng09, Winter09, nonlocalsets10,  SM-2020}. To address the latter direction, the upper bound on locally accessible information, like Holevo bound in global case, was obtained  which is useful to prove local indistinguishability of ensembles with entangled states \cite{HORODECKI-loccdist, HORODECKI-distprotocol} although it fails to capture the results for product ensembles and more nonlocality with less entanglement \cite{Bennett99, Bennett99a,DIVINCENZO-upb,HORODECKI-loccdist}. Some of us have resolved this problem by defining quantumness for ensembles  from two different perspectives -- one is based on the minimal entropy production after dephasing the states in the set of a LOCC distinguishable basis  \cite{HORODECKI-quantumcorrelation} while the other one is based on the generation of entanglement by LOCC indistinguishable sets  of product states under some specific transformations on the whole ensemble \cite{SM-2020}.

In the present work, we characterize quantumness in ensembles consisting of orthogonal product states using measures of coherence \cite{STRELTSOV-coh,BAUMGRATZ-coh}.   In modern day-quantum technology, `coherence' has been shown to be one of the key ingredients which underlies phenomena such as quantum interference \cite{Braun_PRA_2006,Mintert_PRA_2015,MNBera_PRA_2015,Bagan_PRA_2016,Winter_2017}, quantum metrology \cite{Zhang_PRL_2019,RuviLecamwasam_PRX_2024,Ares_OPT_Lett_2021,Silva_PRA_2018}, entanglement \cite{Streltsov_PRL_2015,Chitambar_PRL_2016,Streltsov_PRL_2016,Asmitha_PRAL_2021}, quantum communication \cite{Kelly_SCIPOST_2023,Khan_RSA_2017,Shi_PRA_2017}, thereby establishing it as a resource.  In this respect, see also the recent work which characterizes the coherence of sets \cite{Brunner21}. On the other hand, the quantumness that we want to assess in ensembles using coherence arises from the difficulty in distinguishing states through  \emph{restricted LOCC} protocol. In this context, 'restricted LOCC', referred to as $1$-LOCC, denotes a protocol comprising local operations and a \emph{single round} of classical communication, with the additional requirement that the protocol must succeed regardless of which party initiates it. \textcolor{black}{It is important to highlight that sets of product states classified as indistinguishable under $1$-LOCC are globally orthogonal but can be locally nonorthogonal. This local nonorthogonality within a set of globally orthogonal states, referred to as `the unity of opposites', is inherently tied to the local coherence present in the set with respect to some arbitrary basis.} We uncover how `the unity of opposites' determines local indistinguishability by explicitly constructing coherence-based measure of `quantumness' associated with these sets which is referred to as \textit{minimum ensemble coherence} (\textit{MEC}). Broadly, these sets fall into two categories: $(1)$ those that are deterministically distinguishable by the $1$-LOCC protocol, and $(2)$ those that are probabilistically distinguishable. The measure proposed in this paper provides a quantitative characterization of these distinct classes. Specifically, we prove that $MEC$ vanishes if and only if the complete orthogonal product ensemble is $1$-LOCC distinguishable. Further, we analytically obtain the maximum $MEC$ for the full product basis in qubit-qudit systems. 

When a set of states are not distinguishable by finite  or infinite rounds of LOCC, the natural question is to find their distinguishability probabilistically via LOCC. There are two approaches to implementing imperfect strategies: \emph{unambiguous state discrimination} \cite{CHEFLES-probdist, Hillary05, Duan07}, where the outcomes are always correct, though there are specific probabilities where the protocol may fail to yield an answer; and \emph{minimum error discrimination} \cite{Helstrom_1969, Hayashi08, BARNETT-probdist,VIRMANI200162}, where a conclusion is always reached but with a certain level of error in the success probability, which must be minimized. \textcolor{black}{In this work, we adopt the second strategy and develop it within the framework of the $1$-LOCC protocol. We introduce a coherence-based measure of quantumness for ensembles and demonstrate that this measure vanishes for ensembles that are perfectly distinguishable via $1$-LOCC, indicating an absence of quantumness in such ensembles, while it remains nonzero otherwise. Additionally, we compute the maximum value of this measure achievable by a qubit-qudit ensemble and illustrate the role of mutually unbiased bases (MUBs) in attaining this maximal quantumness. Furthermore, we establish a relationship between the \emph{optimal} success probability of state discrimination for complete product ensembles in two-qubit and qubit-qutrit systems and the coherence-based measure for those ensembles.} 

We organize the paper in the following way. In Sec. \ref{sec:prelim}, we outline the problem and offer an in-depth description of the restricted LOCC protocol, demonstrating the significance of coherence in our work. In Sec. \ref{sec:cohmeasure_completeproductensemble}, we introduce a coherence-based measure for complete product orthogonal ensembles and demonstrate the minimum error discrimination protocol within the restricted LOCC framework. We then establish the effectiveness of this coherence-based measure by connecting it to the optimal success probability in state discrimination. Finally, we conclude in Sec. \ref{sec:conclu}.

\section{Restricted LOCC distinguishability and Coherence measures}
\label{sec:prelim}

We now outline the problem and provide a comprehensive discussion of a restricted local distinguishability scenario, which is relevant to our study. We then present an example to illustrate our objective and explain the relevance of coherence in relation to our work.

\textbf{Formulating the protocol pertinent to our study.} Consider two parties, Alice (\(A\)) and Bob (\(B\)), located at distant sites, who share a state $|\psi_i\rangle_{AB}$, chosen with probability $p_i$, from an ensemble of orthonormal product states, $\{p_i,\ket{\psi_i}_{AB}\in\mathbb{C}^{d_1}\otimes\mathbb{C}^{d_2}\}_{i=1}^{N=d_1\times d_2}$. Their task is to identify the given state using local measurements and a restricted amount of classical communication. Bob measures his part of the system first, communicates the result to Alice, who then performs a measurement on her part based on Bob's outcome, and relays her result to Bob to conclude the protocol. We note that the communication from Alice to Bob is trivial in the sense that no further measurement takes place afterward; its sole purpose is to convey Alice's measurement result. Aside from this trivial communication, the entire protocol is essentially a \emph{one-way}, \emph{single-round} LOCC protocol, denoted as $1\text{-LOCC}_{\text{B}\to\text{A}}$, which indicates that the non-trivial communication flows from Bob to Alice, and it occurs only once. Alice and Bob can also reverse their roles in this scenario, and the corresponding protocol would be denoted as $1\text{-LOCC}_{\text{A}\to\text{B}}$ where the subscript denotes the flow of communication. Together, these protocols are referred to as $1\text{-LOCC}$. A task is said to be perfectly achievable via $1\text{-LOCC}$ if it can be successfully executed using both $1\text{-LOCC}_{\text{A}\to\text{B}}$ and $1\text{-LOCC}_{\text{B}\to\text{A}}$.

\textcolor{black}{\textbf{Formulation of the problem.} The motivation for this task stems from the fact that although Alice and Bob situated in distant locations can create an ensemble, they cannot always distinguish it using communication starting from A as well as B as mentioned above. For example, sometimes an ensemble can be perfectly distinguished via $1\text{-LOCC}_{\text{A}\to\text{B}}$, while $1\text{-LOCC}_{\text{B}\to\text{A}}$ proves inefficient in the same scenario, i.e., a case of \emph{asymmetric distinguishability} \cite{VAIDMAN-distprodst,BARNETT2017-LOCCDISTMINERR}, which again attributed to some kind of quantumness present in the ensemble. This limitation can arise due to the non-orthogonality present in one of the subsystems. It was  demonstrated \cite{Bennett99,Bennett99a} that there exist ensembles composed of orthogonal product states that cannot be perfectly distinguished using \emph{two-way} LOCC (allowing non-trivial communication between A and B in both directions), even with an unlimited amount of classical communication. The structure of these ensembles gives rise to the phenomenon known as \emph{non-locality without entanglement} \cite{Bennett99}. However, in those examples, there is no restriction on the amount of classical communication, unlike in our case. In this work, our focus is on characterizing ensembles that exhibit non-locality (or quantumness) due to the difficulties in distinguishing orthogonal product states under the constraints of the $1\text{-LOCC}$ protocol \footnote{\textcolor{black}{It is relevant to note that in Ref. \cite{BARNETT2017-LOCCDISTMINERR}, the authors explore various levels of complexity in distinguishing orthogonal product states using LOCC. For instance, they discuss scenarios where a set of orthogonal product states can be perfectly distinguished via two-way LOCC, regardless of which party initiates the protocol. Alternatively, some sets can only be perfectly distinguished if a specific party starts the protocol, while others can only be probabilistically distinguished using one-way LOCC.}}. While there have been extensive researches on characterizing states by quantifying resources like entanglement \cite{Horodecki_RMP_2009}, coherence \cite{STRELTSOV-coh}, and other features relevant to information-theoretic tasks, to the best of our knowledge, the investigations to characterize ensembles are limited  in the literature.} 

\textbf{Illustration in two-qubit systems.} We first demonstrate the contrasting features even in the lowest dimension, i.e., in \(\mathcal{C}^{2} \otimes \mathcal{C}^{2}\). Let us consider the computational basis, $\mathcal{E}_1= \{|00\rangle,|01\rangle,|10\rangle,|11\rangle\}$, which is perfectly distinguishable via $1\text{-LOCC}$. Notice that the characteristics of LOCC distinguishability of the ensemble does not change if one replaces \(\{\ket{0}, \ket{1}\}\)  at Alice or Bob's side by \(\{\ket{\eta}, \ket{\eta^{\perp}}\}\), where  \(\ket{\eta} = \cos \frac{\theta}{2} \ket{0} + \exp(i \phi)\sin \frac{\theta}{2}  \ket{1}\) is an arbitrary quantum state and \(\ket{\eta^{\perp}}\) being its corresponding orthogonal state. On the other hand, consider another ensemble of orthogonal product basis, given by  $\mathcal{E}_2= \{|00\rangle,|01\rangle,|1+\rangle,|1-\rangle\}$ \cite{VAIDMAN-distprodst}, where $|\pm\rangle = (\ket{0} \pm \ket{1})/\sqrt{2}$. This ensemble is asymmetrically distinguishable, and can only be distinguished via $1\text{-LOCC}_{\text{A}\to\text{B}}$, while employing $1\text{-LOCC}_{\text{B}\to\text{A}}$ leads to a probabilistic discrimination. If difficulties in local distinguishabillity is a signature of nonclassicality in ensembles, quantumness present in \(\mathcal{E}_2\) is expected to be higher than that of \(\mathcal{E}_1\). Note that Bob's ensemble, \(\{\ket{0},\ket{1},\ket{+},\ket{-}\}\), is used in the well-known Bennett-Brassard (BB84) quantum key distribution protocol \cite{BB84}. In this work, our aim is to capture quantumness present in the two-party ensembles consisting of orthogonal product states in order to characterize them. We believe that if the characterization can capture quantumness present in these non-trivial sets of product ensembles in the lowest dimension, this can be step forward to quantify quantum features in product ensembles. Note further that previous quantification \cite{HORODECKI-loccdist} possibly indicate that LOCC distinguishability for product and entangled ensembles may require different treatments. \textcolor{black}{Among product ensembles, instead of characterizing ensembles which are indistinguishable under unlimited classical communication, we choose an approach by quantifying properties of indistinguishable product ensembles via one-way LOCC with limited amount of classical communication.}

\textbf{Coherence as a good candidate for capturing quantumness.} We now argue that among various quantum properties, coherence stands out as a promising candidate for characterizing ensembles exhibiting non-locality without entanglement within the context of local state discrimination. To this end, notice that, the ensemble $\mathcal{E}_1$ exhibits no local coherence on average with respect to the computational basis, while $\mathcal{E}_2$ displays some degree of local coherence on average. This observation strongly suggests that coherence, indeed, can be a key ingredient in our analysis. 

There exist several coherence quantifiers in the literature  \cite{STRELTSOV-coh}, we here use one of the distance-based coherence measures, namely the \(l_1\) norm of coherence for a state $\rho$ acting on $\mathbb{C}^d$, $C_{l_1}(\rho)$, defined as  
\begin{equation*}
    C_{l_1}(\rho)= \min_{\sigma \in S_I} || \rho - \sigma||_{l_1}=\sum_{i\ne{j}}|{\rho_{ij}}|,
\end{equation*}
where  minimization is taken over the set of incoherent states, \(\sigma\) \cite{BAUMGRATZ-coh}. Since coherence measures depend on the choice of basis, we will primarily use the computational basis (i.e., $\{\ket{i}\}_{i=0}^{d-1}$) as the reference. Any deviation from this will be explicitly mentioned. A state that is diagonal in the reference basis, expressed as $\rho=\sum_{i=0}^{d-1}\rho_{ii}\ket{i}\bra{i}$, will be referred to as incoherent. In contrast, a state that is an equal superposition of the basis states, $\rho=\frac{1}{d}\sum_{i,j=0}^{d-1} e^{(\delta_i-\delta_j)}\ket{i}\bra{j}$, possesses maximal coherence, which is $(d-1)$. \textcolor{black}{Note that while alternative coherence measures could be considered --- such as the relative entropy of coherence, \( C_{\text{rel-ent}}(\rho) = S(\rho_{\text{diag}}) - S(\rho) \), where \( \rho_{\text{diag}} \) denotes the diagonal part of \( \rho \), and $S(\rho) = - \text{tr} (\rho log_2 \rho)$ being the von Neumann entropy --- the main conclusions remain unaffected.}

A key objective of this work is to explore an application of the quantumness inherent in an ensemble. In this context, we can turn to the quantum random access codes (QRAC) strategy, where the effectiveness of encoding can be linked to the coherence-based measure of the ensemble, which will be discussed in the following section (see Remark $2$ in Sec. \ref{subsubsec:psucc_22}).

\section{Coherence-based measures for complete orthogonal product ensembles}
\label{sec:cohmeasure_completeproductensemble}
We first introduce here a coherence-based measure, referred to as \emph{minimum ensemble coherence}, to characterize the quantumness of a \emph{complete orthogonal product ensemble} consisting of a complete set of orthonormal product basis states. We then demonstrate its effectiveness for two-qubit and qubit-qutrit product bases. To begin with, let us first consider a generic complete orthogonal product ensemble of a bipartite system, $\mathcal{E}^{CPB} = \{p_i, |\psi_i\rangle \otimes|\phi_i\rangle\}_{i=1}^{N=d_1\times d_2}$ in dimension $\mathcal{C}^{d_1}\otimes \mathcal{C}^{d_2}$ (henceforth denoted as \(d_1 \otimes d_2\)).\\

\textbf{Definition $1$.} The minimum ensemble coherence ($MEC$) of the ensemble  $\mathcal{E}^{CPB}$ is defined as 
\begin{equation}
\label{eq:MEC}
    MEC(\mathcal{E}^{CPB}) = \underset{\{U_1, U_2\}}{\min} \sum_{i} p_i C_{l_1}\left(U_1\otimes U_2 |\psi_i\rangle \otimes|\phi_i\rangle\right),
\end{equation}
where the minimization is taken over the set of local unitary operators, \(\{U_1, U_2\}\), applied locally on Alice's and Bob's sides. 

\textbf{Theorem $1$.} $\mathcal{E}^{CPB}$ is perfectly distinguishable via $1\text{-LOCC}$, i.e., $\mathcal{E}^{CPB}$ possesses no quantumness, if and only if $MEC(\mathcal{E}^{CPB})$ vanishes.

\textit{Proof.} If $\mathcal{E}^{CPB}$ is perfectly distinguishable, both Alice and Bob possess a single set of basis states each. The application of unitaries $U_1$ and $U_2$ transforms these local bases into the corresponding computational bases. Therefore, Eq. (\ref{eq:MEC}) results in a zero value for $MEC(\mathcal{E}^{CPB})$. Conversely, if $MEC(\mathcal{E}^{CPB})=0$, it indicates that the individual coherence measures $C_{l_1}\left(U_1\otimes U_2 |\psi_i\rangle \otimes|\phi_i\rangle\right)$ become zero after the minimization (as coherence measures cannot be negative). This implies that, following the application of the minimizing unitaries, Alice and Bob obtain a complete product ensemble of computational basis states $\{\ket{i}\otimes\ket{j}\}$, which is perfectly distinguishable via $1\text{-LOCC}$. This confirms that $\mathcal{E}$ is perfectly distinguishable, as local unitaries cannot enhance local distinguishability. ~$\hfill\blacksquare$

\textcolor{black}{\textbf{Remark $1$.} The definition of $MEC$ clearly shows that it remains invariant under the action of local unitary operations. Specifically, we have $MEC(\mathcal{E}^{CPB})\equiv MEC(U_1\otimes U_2\mathcal{E}^{CPB})$. Given that our study focuses solely on ensembles of complete orthogonal pure product states, the behavior of the $MEC$ under general local completely positive trace-preserving (CPTP) operations is unclear at this moment.}

\textcolor{black}{\textbf{Theorem $2$.} The maximum $MEC$ that can be achieved by a $2\otimes d$ ensemble is given by $\frac{d-1}{2}$.}

\textcolor{black}{\textit{Proof.} Without loss of generality, we can write any complete orthogonal product ensemble in $2\otimes d$ as $\mathcal{E}_{2d}^{CPB}=\{p_{i1},p_{i2};\ket{\phi\eta_1^{(i)}},\ket{\phi^\perp\eta_2^{(i)}}\}_{i=1}^d$, where the set $\{\ket{\eta_j^{(1)}},\ket{\eta_j^{(2)}},\cdots,\ket{\eta_j^{(d)}}\}$ forms a complete orthonormal basis for $j=1,2$ and \(p_{ij}\) is the corresponding probability to choose the state from the ensemble. Given that the $l_1$ norm of coherence is super-additive \cite{Yang_2022}, it is evident that the optimal unitary operation on Alice's side should rotate the states $\{\ket{\phi},\ket{\phi^\perp}\}$ to the computational basis states $\{\ket{0},\ket{1}\}$. Thus, using Eq. (\ref{eq:MEC}), we can express 
$$MEC(\mathcal{E}_{2d}^{CPB})=\min_{U_2}\sum_{i,j=1}^{d,2}p_{ij}C_{l_1}(U_2\ket{\eta_j^{(i)}}),$$ where we  utilize the property that $C_{l_1}(\rho\otimes\sigma_I)=C_{l_1}(\rho)$ for any state $\rho$, with $\sigma_I$ being an incoherent state. Therefore, in this scenario, the maximum value of $MEC$ across all possible $2\otimes d$ ensembles is attained when the optimal $U_2$ rotates the basis states $\{\ket{\eta_j^{(i)}}\}_{i=1}^d$ to the computational basis for some specific $j$, while for other value of $j$, the corresponding rotated basis consists of maximally coherent states for each $i$. This leads to the maximum value of $MEC$ being calculated as $\frac{d\times 0 + d\times (d-1)}{2d}=\frac{(d-1)}{2}$. Hence the proof. ~$\hfill\blacksquare$
}\\

\textcolor{black}{\textbf{Example of $2\otimes d$ ensembles achieving maximum $MEC$.} Consider two arbitrary mutually unbiased bases (MUBs) in a Hilbert space of dimension $\mathbb{C}^d$, denoted as $\{\ket{\phi^{(i)}}\}_{i=1}^d$ and $\{\ket{\psi^{(i)}}\}_{i=1}^d$, where $|\braket{\phi^{(i)}|\psi^{(j)}}|=\frac{1}{\sqrt{d}}~\forall {i,j}$. Define $\ket{\chi^{(i)}}=\tilde{U}_{\chi}\ket{i}$ with $\chi\in\{\phi,\psi\}$, where $\{\ket{i}\}$ represents the computational basis. Two notable properties of MUBs are: $(i)$ MUBs remain mutually unbiased under unitary transformations, and $(ii)$ all states from a given MUB exhibit maximal coherence when measured with respect to any other mutually unbiased basis. Now, consider the ensemble $\mathcal{E}_{2d}^{MUB}=\{\frac{1}{2d};\ket{0\phi^{(i)}},\ket{1\psi^{(i)}}\}_{i=1}^d$ in $2\otimes d$. Assume there exists a pair of MUBs, $\{\ket{\phi^{(i)}}\}_{i=1}^d$ and $\{\ket{\psi^{(i)}}\}_{i=1}^d$, such that the optimal unitary $U_2$ in Eq. (\ref{eq:MEC}) becomes either $\tilde{U}_{\phi}^\dagger$ or $\tilde{U}_{\psi}^\dagger$. Leveraging the two aforementioned properties of MUBs, it follows that $MEC(\mathcal{E}_{2d}^{MUB})=\frac{d-1}{2}$, which is the maximal value for $2\otimes d$. In the following subsection, we present explicit examples of ensembles that achieve this maximal $MEC$.}

\textbf{Remark $2$.} Theorem $1$, combined with Remark $1$, provides an intuitive justification for the measure introduced in Definition $1$, establishing it as a meaningful quantifier of the quantumness in an ensemble. Furthermore, we have shown that $MEC(\mathcal{E}^{CPB})$ reaches its maximum value for ensembles where the probability of state discrimination via $1$-LOCC protocol is minimal, highlighting the maximal quantumness arising from the difficulty in state distinguishability (see Figs. \ref{fig:MEC_Psucc_22} and \ref{fig:MEC_Psucc_23}), further reinforcing our justification.

\subsection{$MEC$ for arbitrary product basis in two qubits and its connection with state distinguishability}
\label{subsec:MEC_psucc_22}
\textcolor{black}{Here, we will find the compact form of $MEC$ for arbitrary $2\otimes 2$ ensemble, $\mathcal{E}_{4}^{CPB}$, and connect it with the success probability of state distinguishability in the regime of minimum error state discrimination strategy.}
\subsubsection{$MEC$ for arbitrary two-qubit product ensembles} 
\label{subsubsec:MEC_22}
Let us consider an arbitrary full orthogonal product basis in $2\otimes 2$, given by $\{p_i,\ket{\psi}_i\otimes\ket{\phi}_i\}_{i=1}^4$. Without loss of generality, one can consider a generic complete product ensemble in $2\otimes2$, under the constraint of $1\text{-LOCC}_{\text{B}\to\text{A}}$, represented as $\mathcal{E}_{4}^{CPB}= \{\frac{1}{4};\ket{0\eta_1},\ket{0\eta_1^\perp},\ket{1\eta_2},\ket{1\eta_2^\perp}\}$ where  $\ket{\eta_1}$ and $\ket{\eta_2}$ are arbitrary non-orthogonal qubits, given by  $\ket{\eta_i}=\cos (\theta_i/2) \ket{0} + e^{\iota\alpha_i}\sin (\theta_i/2)\ket{1}$ with $\iota=\sqrt{-1}$. Let us now determine the optimal $U_1$ and $U_2$ to calculate $MEC(\mathcal{E}_{4}^{CPB})$. Since Alice's states are already in the computational basis, the optimal unitary for her is the identity operator. Therefore, our goal is to find the optimal $U_2$, which will allow us to compute $MEC(\mathcal{E}_{4}^{CPB})$ as described in Eq. (\ref{eq:MEC}). To this end, consider the general expression of a two-dimensional unitary,
\begin{equation}
    U_2 = e^{\iota\gamma}
    \begin{pmatrix}
    e^{\iota\beta_1}\cos{(\kappa/2)} & e^{\iota\beta_2}\sin{(\kappa/2)} \\
    -e^{-\iota\beta_2}\sin{(\kappa/2)} & e^{-\iota\beta_1}\cos{(\kappa/2)}
    \end{pmatrix},
    \label{eq:2d_unitary}
\end{equation}
where $0\leq\gamma,\beta_i,2\kappa\leq2\pi$. In our analysis, $\gamma$ is irrelevant throughout the calculations, as it represents a mere global phase factor. Therefore, without loss of generality, we set $\gamma=0$ for the rest of the calculations. Now, the task is to optimize over $\beta_i$s, and $\kappa$ to obtain $MEC$ for a given parameters of the ensemble, $\mathcal{E}_{4}^{CPB}$, i.e., $\{\theta_i, \alpha_i\}_{i=1,2}$. Utilizing the property of the $l_1$ norm of coherence,  $C_{l_1}(\ket{i}\bra{i}\otimes\rho)=C_{l_1}(\rho)$, where $\{\ket{i}\}$ is the basis in which coherence is measured, and considering that the states in the ensemble are equally probable, we find that the optimal $U_2$ occurs at $\kappa=\pi-\theta_1$ and $(\beta_1-\beta_2)=\alpha_1-\pi$, or at $\kappa=\pi-\theta_2$ and $(\beta_1-\beta_2)=\alpha_2-\pi$. Furthermore, it can be verified that both minima yield the same value of $MEC(\mathcal{E}_{4}^{CPB})$. Therefore, using Eq. (\ref{eq:MEC}), we arrive at 
\begin{eqnarray}
\label{eq:MEC_2qubit}
   \hspace{-2em}&&MEC(\mathcal{E}_{4}^{CPB})\nonumber\\\hspace{-2em}&&=\frac{1}{2}\sqrt{1-(\cos\theta_1\cos\theta_2+\cos\alpha\sin\theta_1\sin\theta_2)^2},
\end{eqnarray}
where $\alpha=\alpha_1-\alpha_2$. It is evident that if $\theta_1=\theta_2$ and $\alpha_1=\alpha_2$, implying the ensemble to be perfectly distinguishable by $1\text{-LOCC}_{\text{B}\to\text{A}}$, $MEC(\mathcal{E}_{4}^{CPB})=0$, thereby confirming Theorem $1$. Moreover, the optimal unitary operation to obtain $MEC(\mathcal{E}_{4}^{CPB})$ rotates Bob's states in such a way that one of the bases (e.g., $\{\ket{\eta_1}, \ket{\eta_2}\}$) aligns with the computational basis. Consequently, the maximum achievable $MEC$ is $\frac{1}{2}$, which can also be confirmed from Eq. (\ref{eq:MEC_2qubit}). A prime example of an ensemble achieving maximal $MEC$ in $2\otimes2$ is $\mathcal{E}_4^{MUB}= \{|00\rangle,|01\rangle,|1+\rangle,|1-\rangle\}$. 

Now, we demonstrate that the coherence-based measure, $MEC$, effectively captures the quantumness present in the ensemble $\mathcal{E}_{4}^{CPB}$. To this end, note that if the states in $\mathcal{E}_{4}^{CPB}$ are equally probable, they can be probabilistically distinguished via $1\text{-LOCC}_{\text{B}\to\text{A}}$ using a method similar to that introduced in Ref. \cite{BARNETT2017-LOCCDISTMINERR}.
\subsubsection{Distinguishing states in $\mathcal{E}_{4}^{CPB}$ probabilistically via $1\text{-LOCC}_{\text{B}\to\text{A}}$} 
\label{subsubsec:psucc_22}
\begin{figure}[t] 
\includegraphics[width=8.5cm, height =5cm]{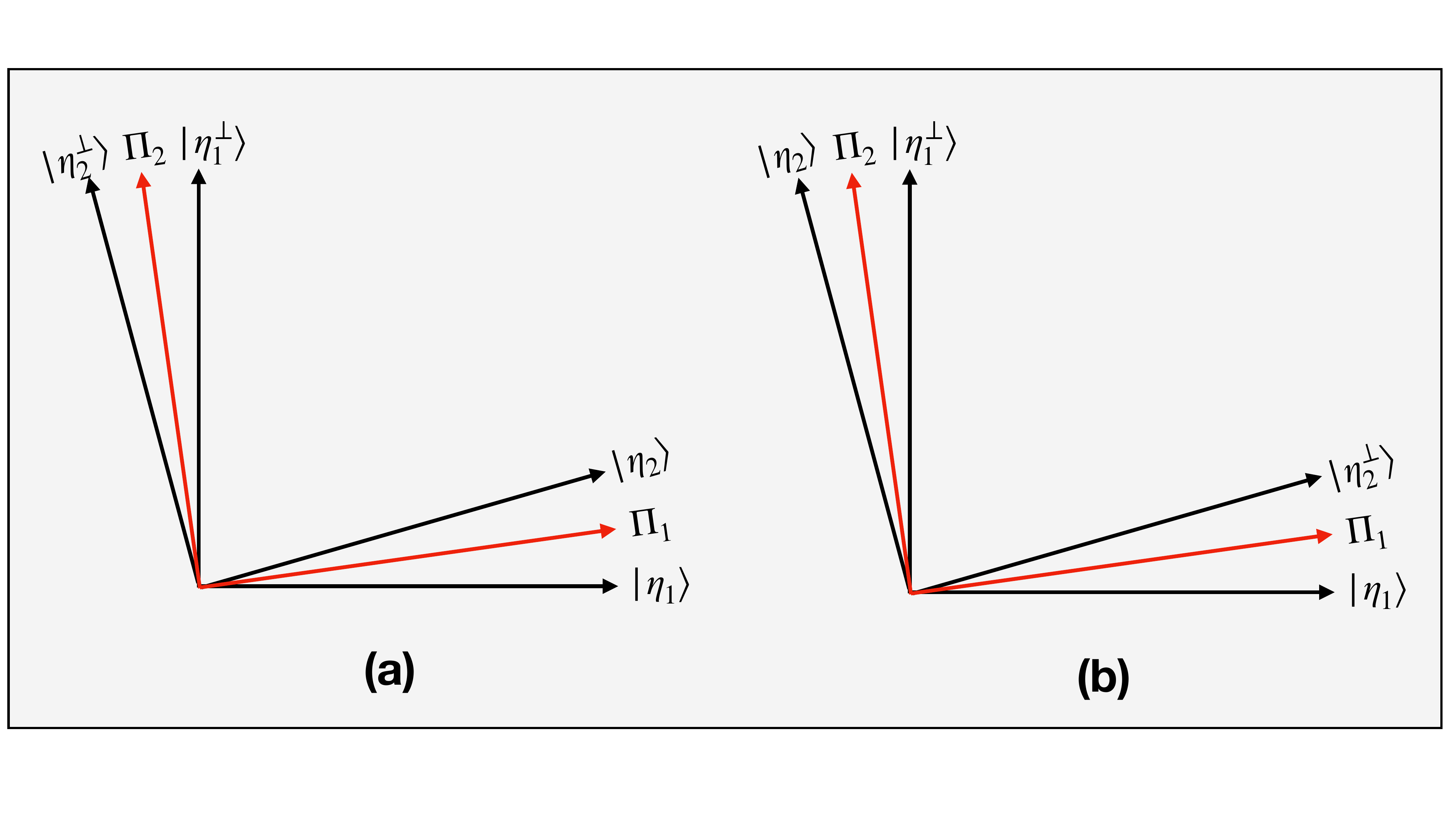}
\caption{Schematic diagram of the possible configurations of the ensemble $\mathcal{E}_{4}^{CPB}$. For a given ensemble in $2\otimes2$ dimension, there are two possible configurations which are shown in (a) and (b). Relevant rank-$1$ projectors to reduce the ensemble to any of the four possible sets, \(S_i, i =1 \ldots 4\), given in Eq. (\ref{eq:config}), are also shown for each configurations.  For case (a), the outcome of the projectors $\Pi_1$ and $\Pi_2$ ensure that the ensemble reduces to the set $S_1$ and $S_2$ respectively, while  for  case (b), the outcome of the same conclude the reduction to $S_3$ and $S_4$ respectively.}
\label{fig:scheme2x2}
\end{figure}

The protocol begins with Bob, who first measures his qubit and communicates the result to Alice via classical communication. Alice then measures her qubit in the $\{\ket{0},\ket{1}\}$ basis and identifies the state that was provided to them from the ensemble. Note that Alice's measurement can distinguish the state if the measurement of Bob reduces the ensemble to one of the four possible sets, given by
\begin{eqnarray}
    \hspace{-1em}S_1&=&\{\ket{0\eta_1},\ket{1\eta_2}\},\; S_2=\{\ket{0\eta_1^\perp},\ket{1\eta_2^\perp}\}, \nonumber \\
    \hspace{-1em}S_3&=&\{\ket{0\eta_1},\ket{1\eta_2^\perp}\}, \; \text{ and } S_4=\{\ket{0\eta_1^\perp},\ket{1\eta_2}\}.
    \label{eq:config}
\end{eqnarray}
To reduce the ensemble into any of the sets $S_i$, Bob can perform a projective measurement $\Pi=\{\Pi_1,\Pi_2\}$, i.e., $\Pi_i\Pi_j=\Pi_i\delta_{ij}$ and $\Pi_1+\Pi_2=\mathbb{I}_2$. If $\Pi_1$ clicks, the ensemble reduces to either $S_1$ or $S_3$, depending on the configuration, as depicted in Fig. \ref{fig:scheme2x2}(a). On the other hand, the ensemble can be either $S_2$ or $S_4$, upon the event of clicking $\Pi_2$. For the configuration in Fig. \ref{fig:scheme2x2}(a), the optimal success probability is given by  $\max\limits_{\{\Pi_1,\Pi_2\}}\frac{1}{4}\sum_{i=1}^2\text{Tr}\left(\Pi_1\ket{\eta_i}\bra{\eta_i}+\Pi_2\ket{\eta_i^\perp}\bra{\eta_i^\perp}\right)=\max\limits_{\{\Pi_1\}}\frac{1}{2}\sum_{i=1}^2\text{Tr}\left(\Pi_1\ket{\eta_i}\bra{\eta_i}\right)=\frac{1}{2}\sum_{i=1}^2\text{Tr}\left(\Pi_{1,\text{opt}}\ket{\eta_i}\bra{\eta_i}\right)$. For an arbitrary ensemble, $\mathcal{E}_{4}^{CPB}$, which can be in either the configuration of Fig. \ref{fig:scheme2x2}(a) or (b), we obtain that 
\begin{eqnarray}
\label{eq:p_succ}
   &&\hspace{-3.8em}\mathcal{P}_{succ}(\mathcal{E}_{4}^{CPB})\max\{p_a,p_b\},~\text{where}\nonumber\\
   &&\hspace{-3.8em}p_{a(b)}=\frac{1}{2}\left(1+\sqrt{\frac{1\pm(\cos\theta_1\cos\theta_2+\cos\alpha\sin\theta_1\sin\theta_2)}{2}}\right).\nonumber\\
\end{eqnarray}
\textcolor{black}{See  \ref{app:psucc_opt} for the detailed analysis of the calculation of success probability, where we have also discussed the optimality of our protocol using the Helstrom bound of state discrimination \cite{Helstrom_1969}.}

\textbf{Remark $3$.} If we minimize the probability, $\mathcal{P}_{succ}$, over the ensemble parameters, the minimum value turns out to be $\frac{1}{2}\left(1+\frac{1}{\sqrt{2}}\right)\approx 0.8535$. Interestingly, in $2\text{-bits}\to 1\text{-qubit}$ (in short, $2^{(2)}\to 1$) quantum random access codes (QRAC), the optimal success probability on average achieves the same value. In fact, one can demonstrate a complementary relationship between these two scenarios: the set of states utilized in optimal QRAC, i.e.,  $\Xi_1=\{\ket{0}$,$\ket{1}$,$\ket{+},\ket{-}\}$ (see Lemma $3.1$ of Ref. \cite{Ambainis_1999}), yield the lowest distinguishing probability under $1$-LOCC, whereas $\{\ket{\eta_1},~\ket{\eta_2}\}\in\text{computational state}$ maximizes distinguishability but do not provide any quantum advantage in the random access codes scheme. In a more general QRAC scenario, where the encoding is performed with the set of states $\Xi_2=\{\ket{\eta_1},\ket{\eta_1^\perp},\ket{\eta_2},\ket{\eta_2^\perp}\}$, with one of the bases $\{\ket{\eta_i},\ket{\eta_i^\perp}\}_{i=1,2}$ being computational, the performance can be attributed to the coherence-based measure, $MEC$, of the ensemble $\mathcal{E}_{4}^{CPB}$. \textcolor{black}{Furthermore, it is intriguing to investigate whether the key rate in the BB84 protocol, achieved using the set $\Xi_2$ (instead of $\Xi_1$), has any connection to the $MEC$ of the ensemble $\mathcal{E}_{4}^{CPB}$.}

\begin{figure}[t] 
\includegraphics[width=1\linewidth]{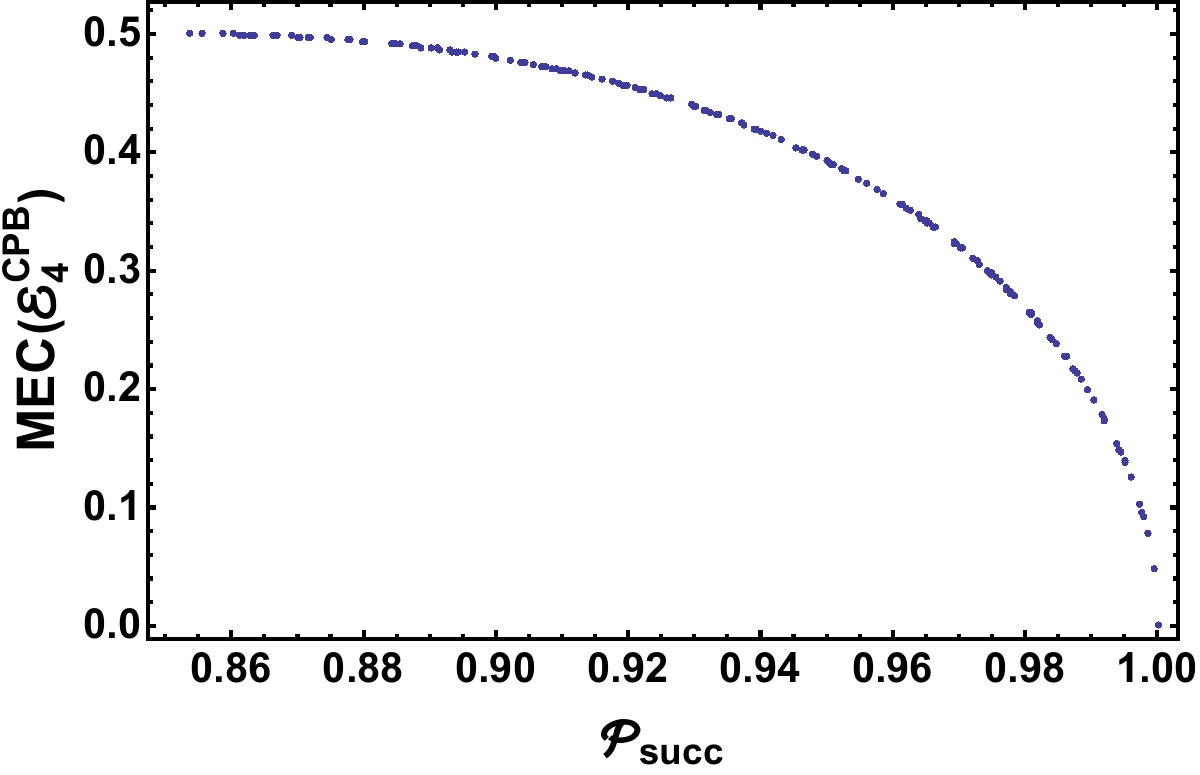}
\caption{
$MEC(\mathcal{E}_{4}^{CPB})$ (ordinate) with respect to \(\mathcal{P}_{succ}\) (abscissa) for arbitrary complete orthogonal product ensembles in \(2 \otimes 2\). Here we generate arbitrary $\mathcal{E}_{4}^{CPB}$ by choosing ensemble parameters (i.e., $\{\theta_i,\alpha_i\}_{i=1}^2$) iteratively in the range $0\leq\theta_i\leq\pi$ and $0\leq\alpha_i\leq2\pi$. The functional relationship between $MEC$ and $\mathcal{P}_{succ}$ is given by Eq. (\ref{eq:MEC_Psucc_22}). Both the axes are dimensionless.
}
\label{fig:MEC_Psucc_22}
\end{figure}

\textbf{Connecting the optimal success probability with $MEC(\mathcal{E}_{4}^{CPB})$.} We are now in a position to explore the relation between $\mathcal{P}_{succ}$ and $MEC$ for the ensemble $\mathcal{E}_{4}^{CPB}$. Using Eqs. (\ref{eq:MEC_2qubit}) and (\ref{eq:p_succ}), we obtain the functional relation between them as
\begin{equation}
\label{eq:MEC_Psucc_22}
    MEC(\mathcal{E}_{4}^{CPB})=2\sqrt{\mathcal{P}_{succ}(1-\mathcal{P}_{succ})(1-2\mathcal{P}_{succ})^2}.
\end{equation}
The above expression suggests that $MEC$ can effectively characterize an ensemble in terms of quantumness it contains, as supported by probabilistic discrimination using $1$-LOCC. In Fig. \ref{fig:MEC_Psucc_22}, we plot $MEC(\mathcal{E}_{4}^{CPB})$ as a function of \(\mathcal{P}_{succ}\) for arbitrary complete orthogonal product ensembles in \(2 \otimes 2\). \textcolor{black}{Here we generate instances of \(\mathcal{E}_4^{CPB}\) by iteratively choosing the ensemble parameters \(\{\theta_i, \alpha_i\}_{i=1}^2\) within the ranges \(0 \leq \theta_i \leq \pi\) and \(0 \leq \alpha_i \leq 2\pi\). Note that the ensembles could be generated using alternative methods, such as by choosing the constituent states from the Haar distribution. However,  we are able to obtain \(MEC(\mathcal{E}_{4}^{CPB})\) in terms of  \(\mathcal{P}_{succ}\) in Eq.~\eqref{eq:MEC_Psucc_22}, and hence, in this case, the resulting plot remains unaffected by the specific ensemble generation procedure.}
The figure shows that a higher $MEC$ indicates greater quantumness, which, in turn, leads to a lower success probability for local state discrimination. \textcolor{black}{Moreover, since the \( l_1 \)-norm and the relative entropy of coherence are monotonically related for pure qubit states, implying that \( MEC(\mathcal{E}_4^{\text{CPB}}) \) computed using the relative entropy of coherence would exhibit the same qualitative features as with the \( l_1 \)-norm, even though the explicit functional dependence on \( \mathcal{P}_{\text{succ}} \) may differ.}

\subsection{$MEC$ beyond two-qubits}
\label{subsec:MEC_psucc_23}
\textcolor{black}{Let us now examine whether the  quantifier for the ensemble extends effectively to higher dimensions. In particular, we will look into the qubit-qutrit ensembles, $\mathcal{E}_{6}^{CPB}$, consisting of complete orthonormal product basis states.}

\subsubsection{$MEC$ for qubit-qutrit ensembles} 
\label{subsubsec:MEC_23}

 Let us consider a generic class of complete orthogonal product ensemble in $2\otimes3$, subject to the constraint of $1\text{-LOCC}_{\text{B}\to\text{A}}$, given by $\mathcal{E}_{6}^{CPB} = \{\ket{0\eta_1},\ket{0\eta_1^\perp},\ket{0\eta_1^{\perp\perp}},\ket{1\eta_2},\ket{1\eta_2^\perp},\ket{1\eta_2^{\perp\perp}}\}$, where $\ket{\eta_i}$, $\ket{\eta_i^{\perp}}$ and $\ket{\eta_i^{\perp\perp}}$ are obtained by applying a general three-dimensional unitary on the computational basis states $\ket{0}$, $\ket{1}$ and $\ket{2}$ respectively.
Once again, in this case, the optimal $U_1$ is the identity operator. The task is to determine the optimal $U_2$, which leads to computation of $MEC(\mathcal{E}_{6}^{CPB})$. To achieve this, consider the general form of a three-dimensional unitary operator,
\begin{eqnarray}
    &\hspace{-3em}U_2=QMQ^T,~\text{where}&\nonumber\\\nonumber\\
    &\hspace{-3em}Q= \begin{pmatrix}
        C_{\kappa} C_{\mu} C_{\chi}+S_{\mu} S_{\chi} & S_{\mu} C_{\chi} - C_{\kappa} C_{\mu} S_{\chi} & C_{\mu} S_{\kappa} \\
        -C_{\kappa} S_{\mu} C_{\chi}+C_{\mu} S_{\chi} & C_{\mu} C_{\chi} + C_{\kappa} S_{\mu} S_{\chi} & -S_{\mu} S_{\kappa} \\
        - C_{\chi} S_{\kappa} & S_{\chi} S_{\kappa} & C_{\kappa}
    \end{pmatrix},&\nonumber\\\nonumber\\
   &\hspace{-3em}\text{and}~M= \begin{pmatrix}
        e^{\iota\beta_1}C_{\zeta} & \iota e^{\iota\beta_2}C_{\nu}S_{\zeta} & \iota e^{\iota\beta_3}S_{\nu}S_{\zeta}\\
        \iota e^{\iota\beta_1}S_{\zeta} & e^{\iota\beta_2}C_{\nu}C_{\zeta} & e^{\iota\beta_3}S_{\nu}C_{\zeta}\\
        0 & e^{\iota\beta_4}S_{\nu} & -e^{\iota(-\beta_2+\beta_3+\beta_4)}C_{\nu}
    \end{pmatrix},
\end{eqnarray}
with $C_x=\cos x$ and $S_x=\sin x$, and $-\pi\leq\mu\leq\pi,~-\pi/2\leq\kappa\leq\pi/2,~0\leq\chi,\beta_i\leq\pi,~-\pi/4\leq\zeta\leq\pi/4,~0\leq\nu\leq\pi/2$. Unlike the two-qubit case, it is not possible to express $MEC(\mathcal{E}_{6}^{CPB})$ in a compact form due to its analytical complexity. However, we can compute $MEC$ numerically and analyze its behavior in relation to the success probability of state discrimination under $1\text{-LOCC}_{\text{B}\to\text{A}}$ for the ensemble $\mathcal{E}_{6}^{CPB}$. Similar to the $2\otimes2$ case, an example of an ensemble achieving maximal $MEC$ in $2\otimes3$, which equals unity, is given by $\mathcal{E}_6^{MUB}=\{|00\rangle,|01\rangle,|02\rangle,|1\bar{\eta}\rangle,|1\bar{\eta}^\perp\rangle,|1\bar{\eta}^{\perp\perp}\rangle\}$, where $\ket{\bar{\eta}}=\frac{1}{\sqrt{3}}\left(\ket{0}+\ket{1}+\ket{2}\right)$, $\ket{\bar{\eta}^\perp}=\frac{1}{\sqrt{3}}\left(\ket{0}+\omega\ket{1}+\omega^2\ket{2}\right)$, $\ket{\bar{\eta}^{\perp\perp}}=\frac{1}{\sqrt{3}}\left(\ket{0}+\omega^2\ket{1}+\omega\ket{2}\right)$, with $\omega=\exp(\frac{2\pi\iota}{3})$.
\subsubsection{Probabilistic state discrimination via $1\text{-LOCC}_{\text{B}\to\text{A}}$}
\label{subsubsec:psucc_23}
Alice and Bob can adopt a similar strategy to the one used in the two-qubit scenario. In order to obtain a conclusive result, the measurement of Bob must reduce the ensemble to one of the nine possible sets, given by
\begin{eqnarray}
\label{eq:conf2x3}
        \hspace{-2.5em}&& S_1=\{\ket{0\eta_1},\ket{1\eta_2}\},  S_2=\{\ket{0\eta_1^\perp},\ket{1\eta_2}\}, \nonumber \\
        \hspace{-2.5em}&& S_3=\{\ket{0\eta_1^{\perp\perp}},\ket{1\eta_2}\}, S_4=\{\ket{0\eta_1},\ket{1\eta_2^\perp}\}, \nonumber  \\
        \hspace{-2.5em}&& S_5=\{\ket{0\eta_1^\perp},\ket{1\eta_2^\perp}\}, S_6=\{\ket{0\eta_1^{\perp\perp}},\ket{1\eta_2^\perp}\}, \nonumber \\
        \hspace{-2.5em}&&S_7=\{\ket{0\eta_1},\ket{1\eta_2^{\perp\perp}}\},S_8=\{\ket{0\eta_1^\perp},\ket{1\eta_2^{\perp\perp}}\},\nonumber \\
        \hspace{-2.5em}&& \text{and } S_9=\{\ket{0\eta_1^{\perp\perp}},\ket{1\eta_2^{\perp\perp}}\}. 
\end{eqnarray}
The six states from the ensemble $\mathcal{E}_{6}^{CPB}$ can be grouped into three sets in six possible ways, resulting in the following six configurations: $\{S_1,S_5,S_9\}$, $\{S_1,S_8,S_6\}$, $\{S_4,S_2,S_9\}$, $\{S_4,S_8,S_3\}$, $\{S_7,S_2,S_6\}$, and  $\{S_7,S_5,S_3\}$. Bob can reduce the ensemble to one of the sets $S_i$ by performing a projective measurement $\Pi=\{\Pi_i\}_{i=1}^3$. When $\Pi_i$ clicks, the ensemble is reduced to one of the sets $\{S_i,S_{i+3},S_{i+6}\}$. For each configuration, $\Pi_j$ selects the set in the $j$-th position. For example, if the ensemble is in the first configuration, $\Pi_1$, $\Pi_2$ and $\Pi_3$ select $S_1$, $S_5$ and $S_9$ respectively, and the optimal success probability by considering only projective measurements is given by $\mathcal{P}_{succ}=\underset{\Pi}{\max}\frac{1}{6}\sum_{i=1}^2\left(\langle\eta_i|\Pi_1|\eta_i\rangle+\langle\eta_i^{\perp}|\Pi_2|\eta_i^{\perp}\rangle+\langle\eta_i^{\perp\perp}|\Pi_3|\eta_i^{\perp\perp}\rangle\right)$. Note that for an arbitrary ensemble $\mathcal{E}_{6}^{CPB}$, the corresponding configuration has to be identified at first, which is possible to determine since the states are known a priory, after which the success probability can be calculated.

\begin{figure}[t] 
\includegraphics[width=1\linewidth]{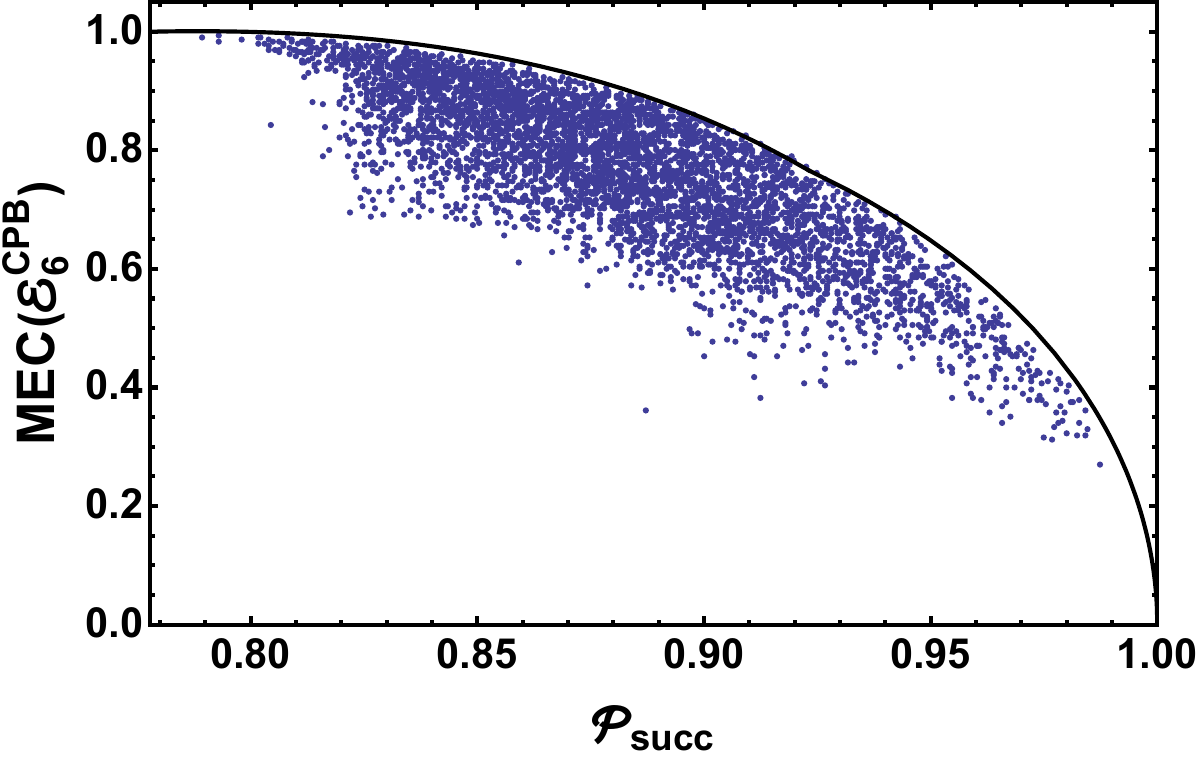}
\caption{
$MEC(\mathcal{E}_{6}^{CPB})$ (ordinate) with respect to \(\mathcal{P}_{succ}\) (abscissa) for arbitrary complete orthogonal product ensembles in \(2 \otimes 3\). Here we  randomly generate $5\times 10^3$ ensembles and numerically compute the upper bound of $MEC$ for a given \(\mathcal{P}_{succ}\). The functional relationship between the upper bound of $MEC$ (the black curve) and $\mathcal{P}_{succ}$ is given by Eq. (\ref{eq:MEC_upperbound_23_psucc}). Both the axes are dimensionless.
}
\label{fig:MEC_Psucc_23}
\end{figure}

\textcolor{black}{\textbf{Note $1$.} Unlike the two-qubit scenario, the protocol for local state discrimination via $1\text{-LOCC}_{\text{B}\to\text{A}}$ described here may not be optimal. Optimality could be achieved by optimizing over a set of all positive operator-valued measures (POVMs) instead of projective measurements, or the lack of optimality might stem from the fact that the discrimination protocol itself is not optimal. \textcolor{black}{However, in the following, we will demonstrate that the connection between $MEC$ and $\mathcal{P}_{succ}$ for ensembles in $2\otimes3$ dimension still holds in the sub-optimal scenario. In particular, we will establish that the upper bound of $MEC$ for a given success probability in $\mathcal{E}_{6}^{CPB}$ follows a trade-off relationship with the success probability, $\mathcal{P}_{succ}$.}}

\textbf{Connecting the probability of success with $MEC(\mathcal{E}_{6}^{CPB})$.} In Fig. \ref{fig:MEC_Psucc_23}, we depict the trend of $MEC$ with respect to the success probability, $\mathcal{P}_{succ}$, for $\mathcal{E}_{6}^{CPB}$. Due to the analytical complexity, we are unable to express $\mathcal{P}_{succ}$ in a compact form. Motivated by the two-qubit scenario, we assume that the ensemble achieving the maximum $MEC$ corresponds to the minimum distinguishing probability. Additionally, we note that no information encoded in a basis $B$ can be recovered by a measurement performed in a basis mutually unbiased to $B$. As previously mentioned, the ensemble $\mathcal{E}_6^{MUB}$ achieves the maximal $MEC$, where $\{\ket{0},\ket{1},\ket{2}\}$ and $\{\ket{\bar{\eta}},\ket{\bar{\eta}^{\perp}},\ket{\bar{\eta}^{\perp\perp}}\}$ are two mutually unbiased bases. Numerical analysis shows that the optimal distinguishing probability for $\mathcal{E}_6^{MUB}$, using projective measurements, is $\frac{7}{9}\approx 0.77778$. Consequently, we infer that the minimal distinguishing probability for ensembles in $2\otimes 3$ systems is $\frac{7}{9}$, corresponding to the ensembles of the form $U_1\otimes U_2\mathcal{E}_6^{MUB}$, where $U_1$ and $U_2$ are arbitrary local unitaries. \textcolor{black}{This inference is further supported by the numerical evidence presented in Fig.~\ref{fig:MEC_Psucc_23}, where we randomly generate \(5 \times 10^3\) ensembles by choosing \(\ket{\eta_1}\) and \(\ket{\eta_2}\) from the Haar uniform distribution (specifically, we generate Haar random qutrit unitary and apply them to the computational basis states). The corresponding values of \(MEC\) are then plotted against \(\mathcal{P}_{succ}\), with the minimum observed success probability being \(0.7891\).} For a given success probability, we numerically compute the upper bound of $MEC$ that a complete orthogonal product ensemble in $2\otimes 3$ can achieve, which takes the form as \footnote{\textcolor{black}{To achieve the upper bound of $MEC$ as given in Eq. (\ref{eq:MEC_upperbound_23_psucc}), we approach it as follows. Drawing inspiration from the $2 \otimes 2$ scenario (specifically Eq. (\ref{eq:MEC_Psucc_22})), we consider the expression within the square brackets in Eq. (\ref{eq:MEC_upperbound_23_psucc}) as a fourth-order polynomial in $\mathcal{P}_{succ}$, with $MEC_{\text{ub}}$ exhibiting symmetry under the exchange of $\mathcal{P}_{succ}$ and $1 - \mathcal{P}_{succ}$. We then use the known values of $MEC_{\text{ub}}$ at $\mathcal{P}_{succ} = \frac{7}{9}$ and $\mathcal{P}_{succ} = 1$, where $MEC_{\text{ub}}$ is $1$ and $0$, respectively. This allows us to parametrize the coefficients of the polynomial by a single parameter. After adjusting this parameter, we arrive at Eq. (\ref{eq:MEC_upperbound_23_psucc}).}}  
\textcolor{black}{
\begin{eqnarray}
  \hspace{-3em}&&~~~~~MEC_{\text{ub}}(\mathcal{P}_{succ})=\max\{f_1(\mathcal{P}_{succ}),f_2(\mathcal{P}_{succ})\}, \nonumber\\
 \hspace{-3em}&&\text{where} \,f_1(x)=\big[\frac{3}{196}x(1-x)(784-2349x(1-x))\big]^{\frac{2}{3}}, \nonumber\\
 \hspace{-3em}&&\text{and} \, f_2(x)=\big[\frac{1}{196}x(1-x)(1960-4779x(1-x))\big]^{\frac{1}{2}}.
 \label{eq:MEC_upperbound_23_psucc}
\end{eqnarray}}
Although, in this case, there is no direct one-to-one correspondence between $MEC$ and $\mathcal{P}_{succ}$ as observed in the two-qubit scenario, the figure shows that as $\mathcal{P}_{succ}$ increases, the upper bound of $MEC$ (i.e., $MEC_{\text{ub}}$, represented by the black curve in Fig. \ref{fig:MEC_Psucc_23}) decreases. \textcolor{black}{ Note also that the numerically obtained upper bound of \(MEC(\mathcal{E}_6^{\text{CPB}})\) may vary depending on the distribution used to generate the ensembles and the coherence measure chosen although the qualitative results do not alter.} This again suggests that, in general, a high value of $MEC$ implies that the product ensembles possess a higher value of quantumness compared to the ones with low $MEC$. Hence, the ensembles with low $MEC$ can be distinguished with high success probability. 

\section{Discussion}
\label{sec:conclu}

In the entanglement resource theory, the free states are the separable ones while the free operations are the local operations and classical communication (LOCC) by which free states can be created. It is natural to predict that the difficulty in discrimination of set of states via LOCC is related to the average entanglement content of the ensembles. However, it was found that such an intuition in LOCC distinguishability does not hold, in general. Specifically, it was surprisingly, reported that there are product ensembles, complete as well as incomplete basis,  which cannot be discriminated by LOCC. 

Characterizing  properties which are responsible for showing LOCC indistinguishability of product as well as entangled ensembles, is one of the central questions in this field.  There have been limited efforts in this area. For instance, the upper bound on locally accessible information, similar to the Holevo bound in the global scenario, has been established. This upper bound is helpful in demonstrating the local indistinguishability of ensembles containing entangled states; however, it does not adequately address results for product ensembles. This issue was partially addressed by examining the entanglement generation from LOCC-indistinguishable sets of product states under specific transformations applied to the entire ensemble. These studies possibly indicate that characterizing product and entangled ensembles has to be done separately. In our work, we quantified quantumness of product ensembles, responsible for LOCC indistinguishability, by using the concept of coherence. 

\textcolor{black}{Specifically, we demonstrated that the average coherence of an ensemble, after appropriately rotating its individual states through unitary operations, can serve as a quantitative measure of the inherent quantumness of product ensembles. This characterization is based on state discrimination using the $1$-LOCC protocol, where $1$-LOCC refers to local operations and a single round of classical communication, with the added condition that the protocol must succeed regardless of which party initiates it. In particular, the coherence-based measure which we call as minimum ensemble coherence ($MEC$) vanishes for those product ensembles which is perfectly distinguishable via $1$-LOCC protocol and is non vanishing otherwise. Furthermore, we established a relationship between the coherence-based measure of an ensemble and the optimal success probability for distinguishing states within that ensemble using $1$-LOCC. For two-qubit product ensembles, we found that $MEC$ can be expressed as a function of the success probability. In higher dimensions, however, this correspondence is not straightforward. Nonetheless, through numerical evidence, we revealed that the relationship is complementary in nature - a higher $MEC$ value corresponds to a lower value of the upper bound on the success probability in the minimum error discrimination protocol. Regarding the applicability of ensemble coherence, we outline a connection between the minimum ensemble coherence and the performance of the quantum random access codes (QRAC) strategy.}


 Among product ensembles, there are several hierarchies present according to their LOCC discrimination protocol. The coherence-based quantifier can capture certain characteristics of ensembles when communication is restricted to a single round. It will be interesting to modify the measure which can capture the hierarchy present in product ensembles according to the rounds of classical communication. Another intriguing direction would be to explore potential measures for ensembles containing both product and entangled states, which could reveal a more intricate structure.

\section*{Acknowledgment}
This research was supported in part by the ``INFOSYS scholarship for senior students''. We acknowledge the support from Interdisciplinary Cyber Physical Systems (ICPS) program of the Department of Science and Technology (DST), India, Grant No.: DST/ICPS/QuST/Theme- 1/2019/23. SM acknowledges the Ministry of Science and Technology, Taiwan (Grant No. MOST 110- 2124-M-002-012) and the National Science and Technology Council, Taiwan (Grants No. 109-2112-M006-010-MY3, 112-2628-M006-007-MY4).  We  acknowledge the use of \href{https://github.com/titaschanda/QIClib}{QIClib} -- a modern C++ library for general purpose quantum information processing and quantum computing (\url{https://titaschanda.github.io/QIClib}) and cluster computing facility at Harish-Chandra Research Institute.

\appendix
\textcolor{black}{
\section{Optimal success probability of state discrimination for the ensemble $\mathcal{E}_4^{CPB}$ via $1$-$LOCC_{B\to A}$}
\label{app:psucc_opt}
Here, we will discuss the optimality of the state discrimination protocol for the ensemble $\mathcal{E}_4^{CPB}$ via $1$-$LOCC_{B\to A}$ as described in Sec. \ref{subsubsec:MEC_22}. Suppose that we are given two pure states $\ket{\psi_1}$ and $\ket{\psi_2}$ with probabilities $q_1$ and $q_2$, respectively, and we have been asked to optimally determine which state is given. To this end, let us briefly recapitulate the Helstrom protocol \cite{Helstrom_1969} for optimal state discrimination. Consider the Hermitian matrix, $\Delta = q_1\ket{\psi_1}\bra{\psi_1} - q_2\ket{\psi_2}\bra{\psi_2}$, which has two eigenstates $\ket{\phi^+}$ and $\ket{\phi^-}$ corresponding to positive and negative eigenvalues. Then the protocol is as follows. The optimal measurement consists of measuring in the basis $\{\ket{\phi^+},\ket{\phi^-}\}$, where obtaining positive (or, negative) eigenstate corresponds to the state $\ket{\psi_1} (\text{or, }\ket{\psi_2})$. The success probability, which is optimal due to Helstrom \cite{Helstrom_1969}, is given as $\mathcal{P}_{succ}\left(\{q_i;\ket{\psi_i}\}_{i=1,2}\right)=1-\left(q_2\text{Tr}\left(\ket{\psi_2}\bra{\psi_2}\ket{\phi^+}\bra{\phi^+}\right)+q_1\text{Tr}\left(\ket{\psi_1}\bra{\psi_1}\ket{\phi^-}\bra{\phi^-}\right)\right)$. Now, consider the ensemble $\mathcal{E}_{4}^{CPB}= \{\frac{1}{4};\ket{0\eta_1},\ket{0\eta_1^\perp},\ket{1\eta_2},\ket{1\eta_2^\perp}\}$ where  $\ket{\eta_1}$ and $\ket{\eta_2}$ are arbitrary non-orthogonal qubits, given by  $\ket{\eta_i}=\cos (\theta_i/2) \ket{0} + e^{\iota\alpha_i}\sin (\theta_i/2)\ket{1}$ with $\iota=\sqrt{-1}$. In Fig. \ref{fig:scheme2x2}, we depict the possible configurations of the ensemble. For the sake of discussion, let us consider that the ensemble is in configuration (a). Using the aforementioned protocol, the optimal success probability to distinguish $\ket{\eta_1}$ and $\ket{\eta_2^\perp}$, respectively, with equal prior probabilities, is found to be 
\begin{eqnarray}
 \hspace{-2.5em}&& \mathcal{P}_{succ}\left(\{\frac{1}{2};\ket{\eta_1},\ket{\eta_2^\perp}\}\right)\nonumber\\  
 \hspace{-2.5em}&& =\frac{1}{2}\left(1+\sqrt{\frac{1+(\cos\theta_1\cos\theta_2+\cos\alpha\sin\theta_1\sin\theta_2)}{2}}\right).\nonumber\\
\hspace{-2.5em} && = p_a.\nonumber
\end{eqnarray}
Similarly, $\mathcal{P}_{succ}\left(\{\frac{1}{2};\ket{\eta_2},\ket{\eta_1^\perp}\}\right)$ turns out to be equal to $p_a$. Further, one can show that the optimal projectors, say, $\{\Pi_{1,\text{opt}},\Pi_{2,\text{opt}}\}$, for both the cases are identical, i.e., obtaining $\Pi_{1,\text{opt}}$ concludes that the state is either $\ket{\eta_1}$ (for $\{\frac{1}{2};\ket{\eta_1},\ket{\eta_2^\perp}\}$) or $\ket{\eta_2}$ (for $\{\frac{1}{2};\ket{\eta_2},\ket{\eta_1^\perp}\}$), while finding $\Pi_{2,\text{opt}}$ infers the state to be either $\ket{\eta_2^\perp}$ (for $\{\frac{1}{2};\ket{\eta_1},\ket{\eta_2^\perp}\}$) or $\ket{\eta_1^\perp}$ (for $\{\frac{1}{2};\ket{\eta_2},\ket{\eta_1^\perp}\}$). 
Considering the state discrimination protocol for the ensemble $\mathcal{E}_4^{CPB}$ via $1$-$LOCC_{B\to A}$ as described in Sec. \ref{subsubsec:MEC_22} when it is in configuration (a), we may note the following inequality:
\begin{eqnarray*}
 \hspace{-3em}&&\max_{\{\Pi_1,\Pi_2\}}\frac{1}{4}\sum_{i=1}^2\text{Tr}\left(\Pi_1\ket{\eta_i}\bra{\eta_i}+\Pi_2\ket{\eta_i^\perp}\bra{\eta_i^\perp}\right)\\
 \hspace{-3em}&&\leq \frac{1}{4}\Big[\max_{\{\Pi_1,\Pi_2\}}\text{Tr}\left(\Pi_1\ket{\eta_1}\bra{\eta_1}+\Pi_2\ket{\eta_2^\perp}\bra{\eta_2^\perp}\right)\\
 \hspace{-3em}&&+\max_{\{\Pi_1,\Pi_2\}}\text{Tr}\left(\Pi_1\ket{\eta_2}\bra{\eta_2}+\Pi_2\ket{\eta_1^\perp}\bra{\eta_1^\perp}\right)\Big]\\
 \hspace{-3em}&&=\frac{1}{2}\left(\mathcal{P}_{succ}\left(\{\frac{1}{2};\ket{\eta_1},\ket{\eta_2^\perp}\}\right)  
  +\mathcal{P}_{succ}\left(\{\frac{1}{2};\ket{\eta_2},\ket{\eta_1^\perp}\}\right)\right)\\
 \hspace{-3em}&&=p_a.
\end{eqnarray*}
Using the fact that the optimal projectors for both the cases being identical, the optimal success probability of state discrimination for the ensemble being in the configuration (a), therefore, is $p_a$. A similar analysis is also performed for the configuration (b), which reveals the corresponding success probability as 
\begin{equation*}
    p_b=\frac{1}{2}\left(1+\sqrt{\frac{1-(\cos\theta_1\cos\theta_2+\cos\alpha\sin\theta_1\sin\theta_2)}{2}}\right).
\end{equation*}
 Hence, for an arbitrary $\mathcal{E}_{4}^{CPB}$, which can be in either the configuration of Fig. \ref{fig:scheme2x2}(a) or (b), we obtain that $\mathcal{P}_{succ}(\mathcal{E}_{4}^{CPB})=\max\{p_a,p_b\}$.}
\bibliographystyle{apsrev4-1}
\bibliography{bib}

\end{document}